\documentclass[reprint, amsmath, amssymb, aps, superscriptaddress
]{revtex4-2}

\usepackage{graphicx}
\usepackage{dcolumn}
\usepackage{bm}
\usepackage{subfigure}
\usepackage{physics}
\usepackage{verbatim}
\usepackage{mathtools}
\usepackage{svg}
\usepackage{hyperref}
\usepackage{color}
\usepackage{soul}

\begin{document}
\preprint{APS/123-QED}

\title{Exploring the neighborhood of 1-layer QAOA\\with Instantaneous Quantum Polynomial circuits}

\author{Sebastian Leontica}
\email{sebastian.leontica.22@ucl.ac.uk}
\affiliation{Quantinuum, Partnership House, Carlisle Place, London SW1P 1BX, United Kingdom}
\affiliation{CMMP Research Group, University College London, Dept of Physics and Astronomy,\\Gower Street, London WC1E 6BT, United Kingdom}
\author{David Amaro}
\email{david.amaro@quantinuum.com}
\affiliation{Quantinuum, Partnership House, Carlisle Place, London SW1P 1BX, United Kingdom}

\date{\today}

\begin{abstract}
We embed 1-layer QAOA circuits into the larger class of parameterized Instantaneous Quantum Polynomial circuits to produce an improved variational quantum algorithm for solving combinatorial optimization problems. 
The use of analytic expressions to find optimal parameters classically makes our protocol robust against barren plateaus and hardware noise. 
The average overlap with the ground state scales as $2^{-0.31(2) N}$ with the number of qubits $N$ for random Sherrington-Kirkpatrick (SK) Hamiltonians of up to 29 qubits, a polynomial improvement over 1-layer QAOA.
Additionally, we observe that performing variational imaginary time evolution on the manifold approximates low-temperature pseudo-Boltzmann states.
Our protocol outperforms 1-layer QAOA on the recently released Quantinuum H2 trapped-ion quantum hardware and emulator, where we obtain an average approximation ratio of $0.985$ across 312 random SK instances of 7 to 32 qubits, from which almost $44\%$ are solved optimally using only 4 to 1208 shots per instance.
\end{abstract}

\maketitle

 \section{Introduction} \label{sec:intro}
Since its introduction by Farhi et al.~\cite{Farhi2014} in 2014, the Quantum Approximate Optimization Algorithm (QAOA) has been explored in the quantum computing literature as one of the most promising heuristics for achieving quantum advantage on near-term devices~\cite{Farhi2015, Jiang2017}. This is only one example of a larger class of variational quantum optimization algorithms, which attempt to produce good solutions to combinatorial optimization problems by sampling a parameterized quantum circuit~\cite{Diez_2021, Liu_2022, Ebadi_2022, Amaro2022, Amaro2022NSC}. 
In the absence of full quantum error correction~\cite{Litinski2019} the required circuits must be sufficiently shallow to withstand noise, yet expressive enough to find states with high overlap onto the ground state. 
QAOA is a particularly good choice for satisfying these criteria, as it has an adjustable number of layers $p$. It can be understood as a Trotterized version of the quantum adiabatic algorithm (QAA), for which compelling theoretical evidence of performance exists~\cite{Farhi2001}.
Additionally, it was shown that even for small numbers of layers, sampling from the QAOA ansatz is a hard task for classical computers~\cite{Farhi2016}.

\begin{figure}[t!]
    \includegraphics[width=0.47\textwidth]{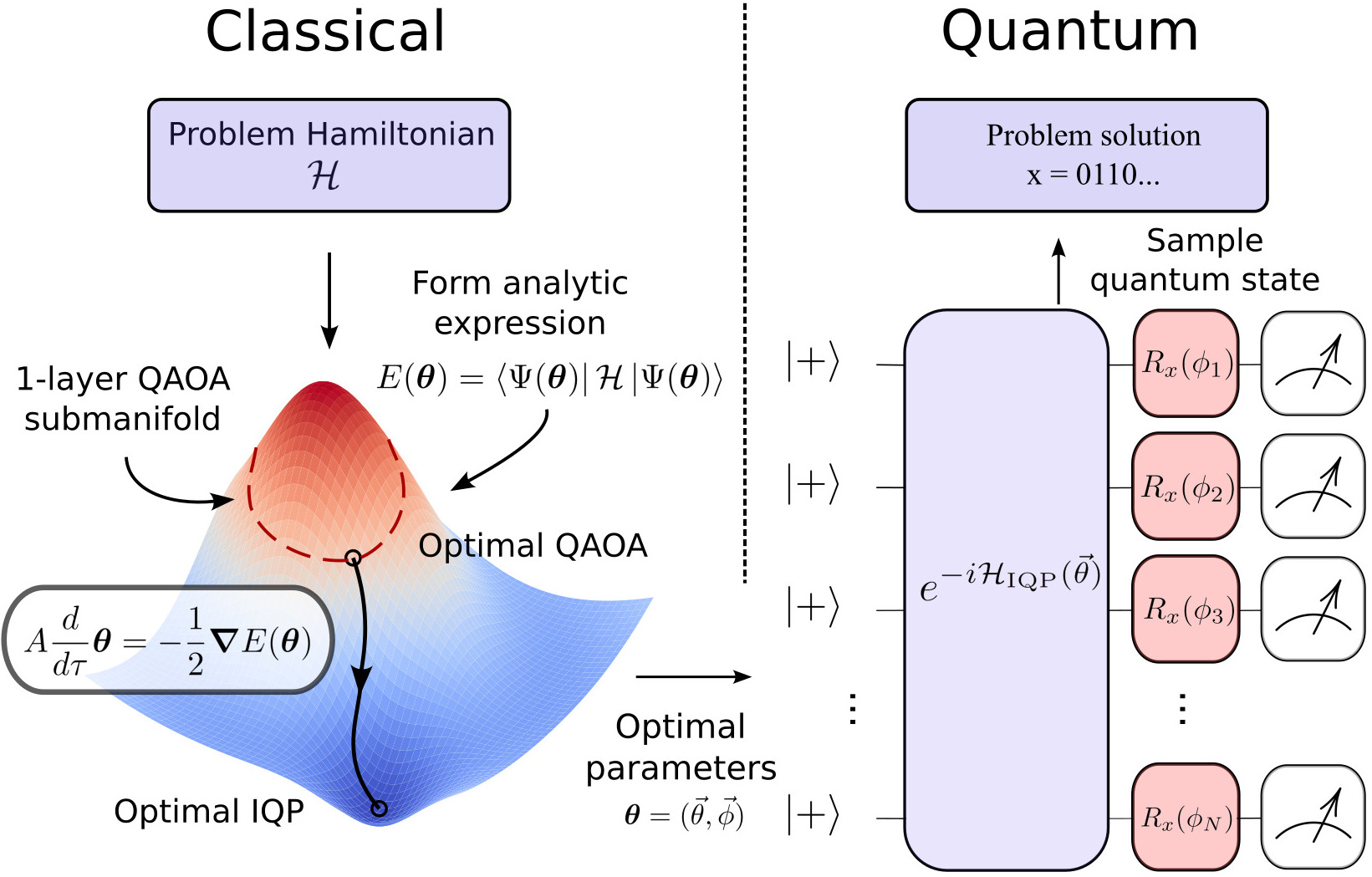}
    \caption{Diagrammatic representation of the algorithm. The 1-layer QAOA ansatz is a submanifold of the IQP ansatz and provides a warm start in the optimization protocol. The trajectory between the QAOA optimum and the IQP optimum is defined via the McLachlan variational principle and is computed classically. Color coding the optimization landscape represents the effective temperature of the associated state, with lower temperature states (blue) having a higher chance of sampling the ground state. The quantum computer is only used during the sampling step, which is known to be difficult classically.}
    \label{fig:landscape}
\end{figure}

In this regime of a small number of layers, the form of the Trotterized QAOA operators may not be the best choice. This has motivated~\cite{Herrman2022, Shi2022, Chalupnik2022,Lotshaw2023} the addition of extra parameters to the QAOA ansatz so that, instead of evolving the state according to the problem Hamiltonian, each parameter in the ansatz has the freedom to evolve independently. By doing this, an ansatz of the same depth may incorporate corrections that would otherwise require multiple layers. 

In particular, 1-layer QAOA circuits--with and without the additional parameterization--belong to the class of parameterized quantum circuits known as Weighted Graph States (WGS) used to simulate condensed matter systems~\cite{Anders2006, Anders2007, Hartmann2007, Plato2008, Hubener2009, Hubener2011, Schindler2022}. For these states, the reduced density matrix in a subsystem of fixed size can be computed classically, allowing the efficient evaluation of local observables on a classical computer. This property permits the derivation of analytic and exact expressions for 1-layer QAOA on arbitrary local Hamiltonians~\cite{Wang2018} and for extra-parameterized circuits on some restricted local Hamiltonians\cite{Herrman2022, Shi2022}. Such expressions are used to train the model classically, bypassing typical limitations such as the appearance of barren plateaus~\cite{Cerezo2021}.

In this manuscript, we explore the embedding of 1-layer QAOA into the broader class of parameterized Instantaneous Quantum Polynomial (IQP) circuits, for which similar hardness of sampling theorems exist~\cite{Shepherd2009, Bremner2016}, even in the presence of moderate noise~\cite{Bremner2017}. IQP circuits also belong to the class of WGS, but compared to QAOA and existing extra-parameterized variants our ansatz uses all-to-all two-qubit interactions, making its implementation problem-independent and most natural for trapped-ion quantum computers. We additionally show that analytic and exact expressions can be obtained for arbitrary local Hamiltonians, and use them to train the model via robust classical techniques like the Runge-Kutta method~\cite{Dormand1980}. We emphasize the role of starting the training from the optimal QAOA and finding a nearby local minimum rather than aiming for a global optimum, which avoids the challenging exploration of non-trivial landscapes~\cite{Bittel2021}.

This leaves only the key ingredient of sampling from the final quantum state to be performed on the quantum device, as illustrated in Fig.~\ref{fig:landscape}. A recent investigation of the states produced by 1-layer QAOA~\cite{Diez-Valle2023} shows that sampling produces a distribution close to a Boltzmann distribution, at temperatures beyond the reach of classical sampling techniques such as Markov Chain Monte Carlo (MCMC)~\cite{Eldan2020}. We improve on this result by lowering the temperature further, using variational quantum imaginary time evolution (VarQITE)~\cite{Yuan2019, McArdle2019}. However, the constraint of keeping the state in the variational manifold limits our ability to follow exact imaginary time evolution, distorting the distribution. 

The manuscript is structured as follows. Section~\ref{sec:qaoa} provides a brief review of QAOA. Our IQP ansatz is introduced in Section~\ref{sec:IQP}, where we make the connection to 1-layer QAOA, describe the derivation of analytical expressions and how to use them for classical training, and discuss a previous work~\cite{Lee2021} that challenges the possibility of quantum advantage with IQP circuits. In Section~\ref{sec:methods} we describe our protocol for approximating thermal distributions and solving combinatorial optimization problems, while Section~\ref{sec:results} presents numerical performance results. First, the average overlap with the ground state obtained with an exact state-vector simulator is polynomially better than for 1-layer QAOA on random Sherrington-Kirkpatrick (SK) Hamiltonians of up to 29 qubits. Second, when approximating thermal distributions we can reach lower temperatures than 1-layer QAOA but the approximation quality reduces. Third, we demonstrate a better performance than 1-layer QAOA at solving random SK Hamiltonians of up to 32 qubits in the recently released Quantinuum's trapped-ion H2 quantum hardware and emulator. Using a reduced number of shots, the best solution per instance presents a large approximation ratio and is optimal for a large fraction of instances. Finally, Section~\ref{sec:discussion} discusses the methods, results, and future research directions.

\section{The Quantum Approximate Optimization Algorithm} \label{sec:qaoa}
The standard implementation of the QAOA~\cite{Farhi2014} attempts to create states with large overlap onto the ground eigenspace of some optimization problem, typically defined through an Ising Hamiltonian,
\begin{equation}
\label{eq:problemHam}
    \mathcal{H} = \sum_i h_i Z_i + \sum_{i<j} J_{ij} Z_i Z_j,
\end{equation}
where the $Z_i$ variables can be interpreted as the projections onto the Z-axis of a classical or quantum mechanical ensemble of $N$ spin-$\frac{1}{2}$ particles and $(h_i, J_{ij})$ are real coefficients. This is achieved by starting with the ground state $\ket{+}^{\otimes N}$ of the trivial transverse field mixing Hamiltonian $\mathcal{H}_x = -\sum_i X_i$ and evolving the state under the alternating application of the propagators of $\mathcal{H}$ and $\mathcal{H}_x$. The final trial state is of the form
\begin{equation}
    \ket{\Psi(\boldsymbol{\gamma},\boldsymbol{\beta})} = \prod_{k=1}^p\exp\left(i\beta_k \mathcal{H}_x\right)\exp\left(-i\gamma_k \mathcal{H}\right) \ket{+}^{\otimes N},
\end{equation}
where $p$ is called the level of the QAOA and the sets $\boldsymbol{\beta}$, $\boldsymbol{\gamma}$ of real coefficients $\beta_k$, $\gamma_k$ are used as variational parameters. The most commonly used cost function in the optimization of the ansatz is the expectation value of the problem Hamiltonian
\begin{equation}
\label{eq:cost}
    E = \bra{\Psi(\boldsymbol{\gamma},\boldsymbol{\beta})}\mathcal{H}\ket{\Psi(\boldsymbol{\gamma},\boldsymbol{\beta})},
\end{equation}
although alternative objective functions have been proposed~\cite{Li2020, Barkoutsos2020}.
For the rest of this work, we will only consider the 1-layer QAOA, which is sufficiently shallow to withstand the effects of moderate noise and obtains an enhanced average probability of sampling the ground state quadratically larger than random guessing~\cite{Diez-Valle2023}, i.e., scaling as $2^{-0.5N}$.

\begin{figure*}[t!]
     \subfigure[]{
         \includegraphics[width=0.485\textwidth]{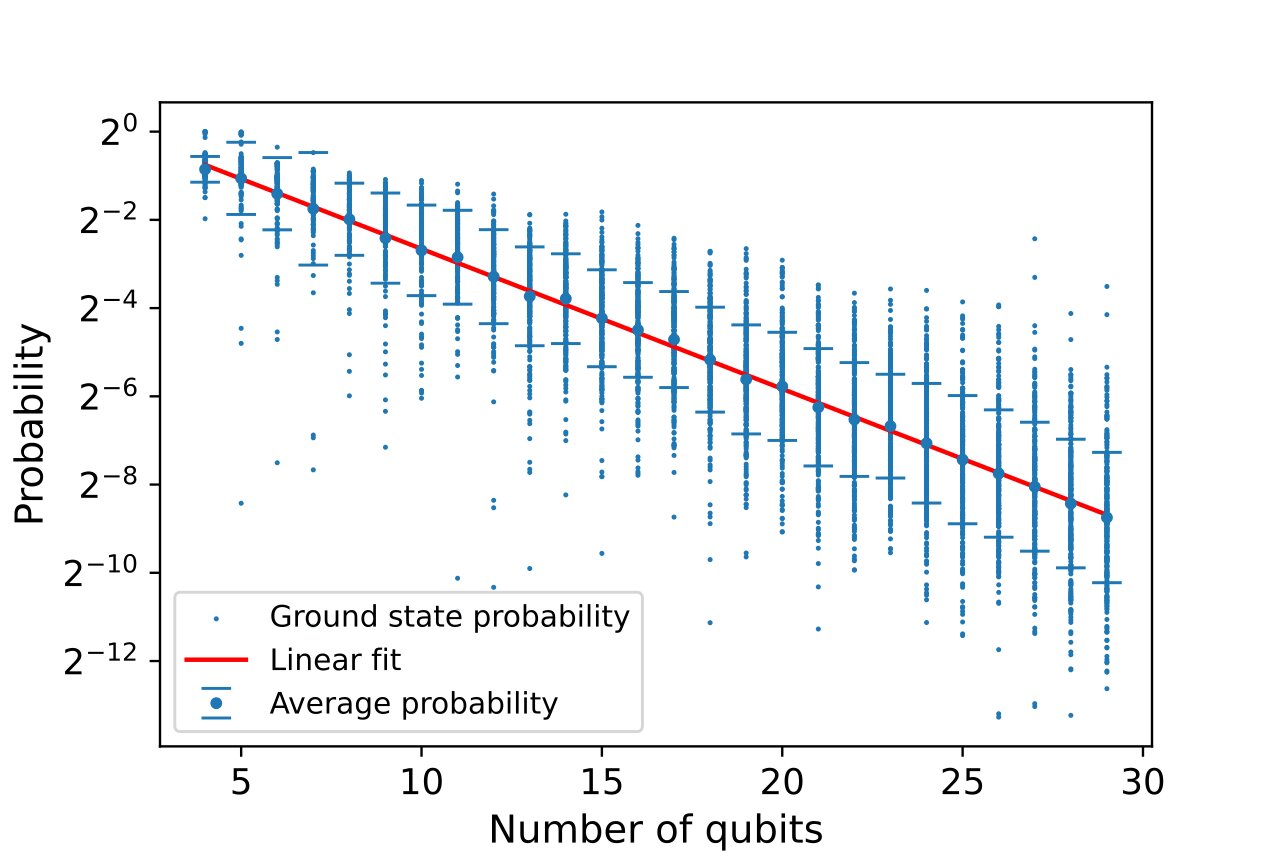}}
     \hfill
     \subfigure[]{
         \includegraphics[width=0.485\textwidth]{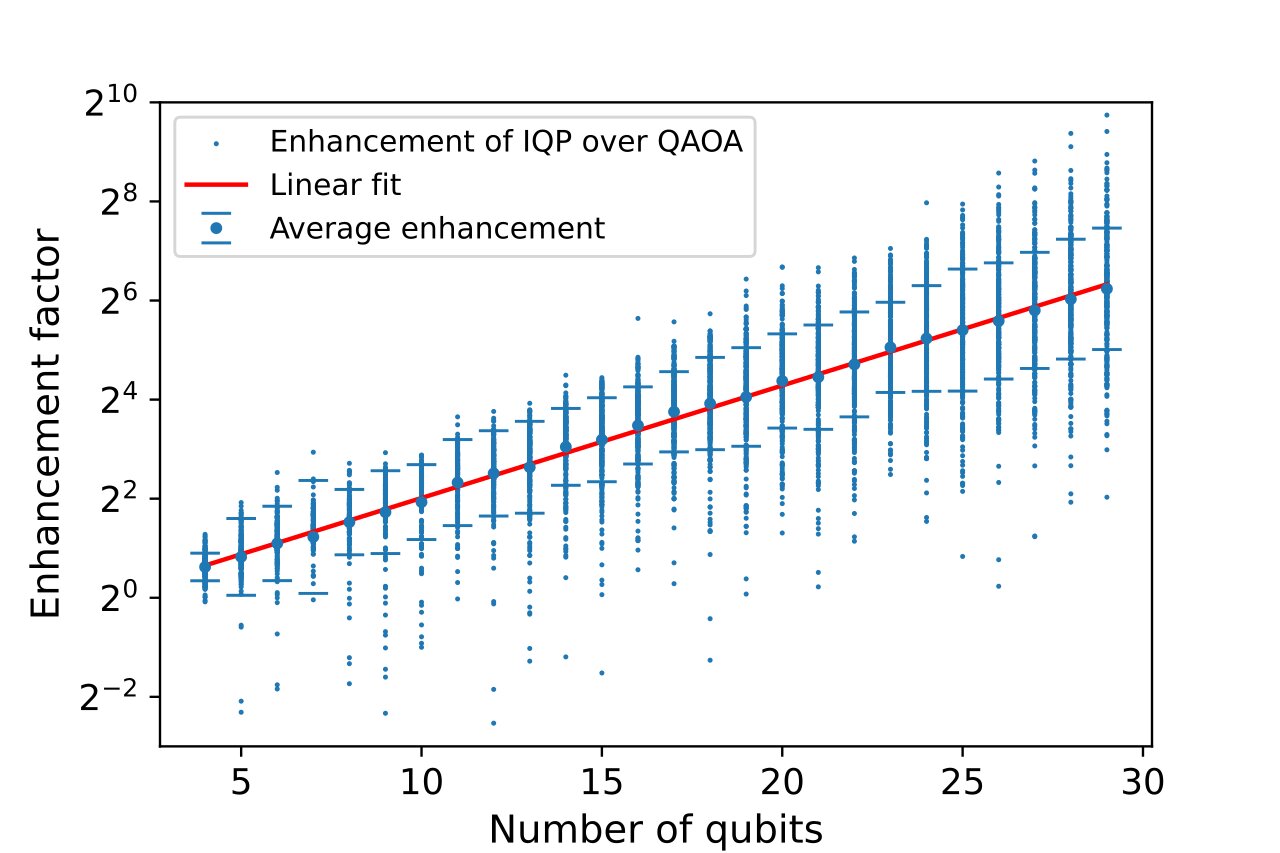}
     }
       \caption{Optimization results for 300 randomly generated Sherrington-Kirkpatrick Hamiltonians of up to 29 spins. a) Probability of sampling the ground state configuration in the optimal IQP ansatz. b) Enhancement factor $p_{\mathrm{IQP}}/p_{\mathrm{QAOA}}$ for finding the ground state in the optimized IQP ansatz compared to the original QAOA. The IQP was optimized until convergence using simple gradient descent. Using a linear fit, we find the average probability of sampling the ground state $p_{\mathrm{IQP}} \sim 2^{-\alpha N}$ with $\alpha = 0.31 \pm 0.02$ and the average enhancement factor $p_{\mathrm{IQP}}/p_{\mathrm{QAOA}} \sim 2^{\delta N}$ with $\delta = 0.23 \pm 0.02$. The errors indicate the variability in gradient at one standard deviation. 
       }
        \label{fig:solver}
\end{figure*}

\section{The Instantaneous Quantum Polynomial circuit} \label{sec:IQP}
The IQP is a non-universal model of quantum computation with similar roots to the boson sampling problem, whose aim is to strengthen the general belief that quantum computers are more powerful than classical machines~\cite{Shepherd2009, Bremner2016}. Under certain widely believed complexity-theoretic assumptions, sampling from the IQP state $H^{\otimes N}\text{exp}(-i\mathcal{H}_{\mathrm{IQP}}(\vec{\theta}))\ket{+}^{\otimes N}$ in the computational basis of all qubits is a hard task for a classical computer~\cite{Bremner2016}. Here the IQP Hamiltonian is defined as $\mathcal{H}_{\mathrm{IQP}}(\vec{\theta}) = \frac{1}{2} \sum_{i} \theta_i Z_i +\frac{1}{2}\sum_{i<j}\theta_{ij} Z_i Z_j$ and $H$ is the Hadamard gate.

The IQP ansatz employed in this work is a generalization where Hadamard gates are replaced with independent parameterized single-qubit rotations $R_x(\phi) = \exp(-i\phi X/2)$, leading to the quantum circuit
\begin{equation} \label{eq:ansatz}
    \ket{\Psi(\boldsymbol{\theta})} = \bigotimes_{i\in \mathcal{N}} R_x(\phi_i) \cdot\exp\left(-i \mathcal{H}_{\mathrm{IQP}}(\vec{\theta})\right) \ket{+}^{\otimes N}, \\
\end{equation}
where $\boldsymbol{\theta} = (\vec{\phi},\vec{\theta})$ are free, real parameters. The IQP state is recovered by setting $\phi_i = \pi/2$ and making the transformation $\theta_i \, \longrightarrow \theta_i - \pi/2$. Since the IQP state can be brought to this form by modifying the final layer of single qubit rotations, we expect generic states of this form to be difficult to sample classically as well. We also make the important observation that, up to single qubit rotations and energy rescaling, the IQP state in Eq.~\eqref{eq:ansatz} is the same as that produced by a 1-layer QAOA designed to solve for the ground state of $\mathcal{H}_{\mathrm{IQP}}$.

This ansatz generalizes the optimization cost function of Eq.~\eqref{eq:cost} to
\begin{equation}
\label{eq:costnew}
    E(\boldsymbol \theta) = \bra{\Psi(\boldsymbol{\theta})} \mathcal{H} \ket{\Psi(\boldsymbol{\theta})} = \langle \mathcal{H}\rangle_{\boldsymbol{\theta}},
\end{equation}
which we refer to as the optimization landscape. The task of computing the cost function defined in Eq.~\eqref{eq:costnew} is then reduced to estimating the expectation values of the spins $\langle Z_i \rangle_{\boldsymbol{\theta}}$ and correlators $\langle Z_i Z_j \rangle_{\boldsymbol{\theta}}$ in an arbitrary state $\ket{\Psi(\boldsymbol{\theta})}$. 
In the Supplemental Material we show that the latter expression can be reduced to calculating partition functions of reduced Ising Hamiltonians of the form
\begin{equation}
\label{eq:partfunc}
    \mathcal{Z}_e = \frac{1}{2^N} \sum_{\{x\}} e^{-i\mathcal{H}_{e}(x)},
\end{equation}
where $e$'s are single or two qubit subsets. The reduced generator $\mathcal{H}_{e}$ retains only the terms in $\mathcal{H}_{\mathrm{IQP}}$ that anti-commute with the operator $X_e = \bigotimes_{i \in e} X_i$. This leads to a highly restricted graph topology, for which partition functions can be evaluated exactly. We generalize this method to show that IQPs have simple analytic expressions for all expectation values of the form $\langle Z_e \rangle_{\boldsymbol{\theta}}$, with a number of terms that scales like $\mathcal{O}(2^{\abs{e}})$. 

These properties of the IQP ansatz make it a good candidate for solving optimization problems, as it is guaranteed to be at least as powerful as 1-layer QAOA and the training can be performed efficiently using only classical resources. The access to exact, analytic expressions for the cost function also means we do not need to worry about finite sampling or device errors during training. Barren plateau issues can also be ruled out, as we can evaluate gradients to arbitrary precision and use adaptive step sizes. 
Access to a quantum computer is only necessary during the final sampling step, so we expect our protocol to perform well under moderate hardware noise.

\begin{figure}[t!]
    \includegraphics[width=0.45\textwidth]{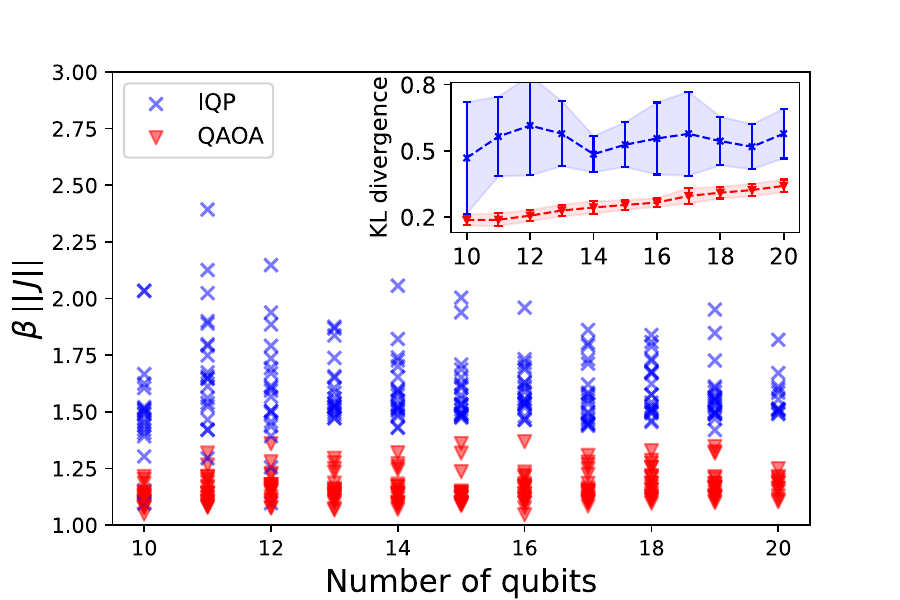}
    \caption{Normalized effective inverse temperatures $\beta \norm{J}$ in the QAOA state and the IQP state after VarQITE evolution for a time $\tau = 10$, for 20 randomly generated Sherrington-Kirkpatrick Hamiltonians of each size from 10 to 20 qubits. We also show the average and standard deviations for the KL divergences of each problem size.}
    \label{fig:temperature}
\end{figure}

As opposed to the standard QAOA ansatz, the IQP is sufficiently flexible to produce all computational states. In particular, this means that, if a classical algorithm were able to find the global optimum of Eq.~\eqref{eq:costnew}, it would also find the exact ground state of $\mathcal{H}$. In~\cite{Lee2021}, it is shown that the optimization landscapes of IQP ansatze with only polynomially many terms (like our ansatz) are generally non-convex and computational states other than the solution may form local minima, which we call trivial minima. Consequently, converging to such local minima would imply the algorithm does not need access to a quantum computer, as the bits $x_i$ of the solution corresponding to the optimal parameters are given by $\langle Z_i \rangle_{\boldsymbol \theta}$, which can be efficiently computed classically.

We prove that the optimization landscapes can contain non-trivial minima, and give a minimal example of this in the Supplemental Material. Remarkably, we provide numerical evidence that for the SK model such a local minimum is located in the vicinity of the QAOA parameters, and show that sampling the IQP circuit at this point greatly enhances the chance of finding the ground state compared to QAOA.

\section{Methods} \label{sec:methods}
A remarkable result of~\cite{Diez-Valle2023} is that for a wide range of optimization problems that can be formulated as in Eq.~\eqref{eq:problemHam}, the 1-layer QAOA is capable of approximating pseudo-Boltzmann states proportional to $\exp(-\beta \mathcal{H}/2)\ket{+}^{\otimes N}$, with large inverse temperature $\beta$, up to relative phases that do not affect the distribution. This is important because sampling this state produces the same distribution as sampling the mixed thermal state $\rho_{\beta} = e^{-\beta \mathcal{H}}/\mathcal{Z}$ for classical Hamiltonians, which is useful for a variety of optimization tasks. 

In our work, we use this result to justify the QAOA as a good starting point in optimizing the IQP ansatz. Since the 1-layer QAOA ansatz can be recovered by restricting the parameters of the full IQP, we find the optimal QAOA position classically, using the BFGS~\cite{Fletcher1987} algorithm on the submanifold. To find a local optimum in the vicinity of this position, it is sufficient to use simple gradient descent. However, we also explore the feasibility of our algorithm for producing low-energy thermal states, which is achieved using a different approach called VarQITE~\cite{Yuan2019, McArdle2019}. This protocol aims to find the trajectory on the manifold that best approximates the action of $\exp(-\tau \mathcal{H})$ on the state. If the initial state is pseudo-Boltzmann, then applying this operator leads to a decrease in temperature. The parameters in the ansatz are evolved according to the McLachlan variational principle~\cite{McLachlan1964}:
\begin{equation}
    A\frac{d}{d\tau} \boldsymbol{\theta} = -\frac{1}{2}\grad E(\boldsymbol{\theta}),
\end{equation}
where the coupling matrix $A$ describes the geometry of the variational manifold (i.e. it is the Gram matrix of the tangent vectors corresponding to each parameter) and $\tau$ is the imaginary time variable. In the Supplemental Material we show that the coefficients of the Gram matrix can be expressed as expectation values of low-weight Pauli operators in the IQP, for which we find simple analytic expressions. However, this calculation is computationally expensive, so when the focus is on finding a local minimum rather than preserving a thermal profile, we set $A = I$ and perform simple gradient descent. 

In both cases, this linear system of ODEs defines a flow on the variational manifold, that we solve numerically using the Runge-Kutta method~\cite{Dormand1980}. We stop this procedure when we arrive at a local minimum, or when $A$ becomes non-invertible. This typically happens after a long plateau in the energy profile, which we illustrate in the Supplemental Material. Such event becomes a rare occurrence when we increase the number of qubits, but for problems that exhibit this behavior, we choose the optimal parameters in the middle of the plateau. After finding the optimal parameters, we sample the circuit and compute the probability of finding the ground state. We share the code used for implementing this protocol in \cite{Leontica2022}.

We characterize our distributions using an effective inverse temperature $\beta$. This is obtained by minimizing the Kullbach-Leibler divergence of the IQP distribution to the family of thermal distributions. Here, we compute the KL divergence exactly, but in practice, this would be estimated from samples~\cite{Benedetti2016}. 

\begin{figure}[t!]
    \includegraphics[width=0.45\textwidth]{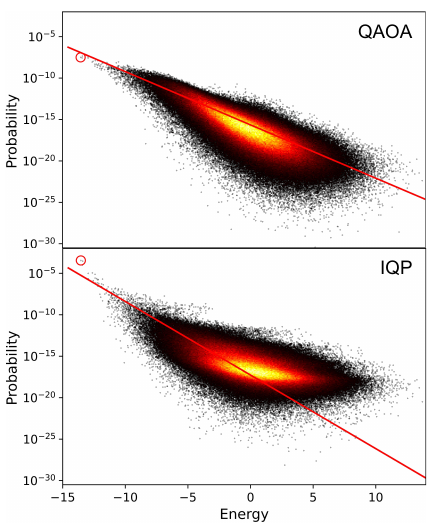}
    \caption{Overlap of the state produced by our ansatz onto different Hamiltonian eigenvalues as a function of energy for the QAOA parameters (top), and optimized IQP parameters (bottom), for a randomly generated 20 qubit Sherrington-Kirkpatrick Hamiltonian. Brighter color indicates higher coarse-grain point density. Red line illustrates the thermal distribution model that minimizes the KL divergence. A red circle marks the location of the ground state. 
    }
        \label{fig:samples}
\end{figure}

\section{Results} \label{sec:results}
We test our method on Sherrington-Kirkpatrick (SK) Hamiltonians~\cite{Sherrington1975, Parisi1980, Panchenko2012} of up to $N=29$ spins using Qiskit exact state-vector quantum simulators~\cite{Qiskit}. These Hamiltonians are of the form of Eq.~\eqref{eq:problemHam} with $h_i = 0$ and $J_{ij}$ independent and identically distributed Gaussian random variables of 0 mean and a standard deviation of $1/\sqrt{N}$. The unbiased SK model presents a $\mathbb{Z}_2$ symmetry, so the ground state is unique up to flipping all qubits. This is a well-understood spin model with compelling classical solvers~\cite{Montanari2019}. In quantum optimization it is one of the most studied benchmark problems~\cite{Sung2020, Google2021, Babbush2021, Dalzell2022, Farhi2022, Diez-Valle2023}. 

\begin{figure*}[t!]
    \includegraphics[width=\textwidth]{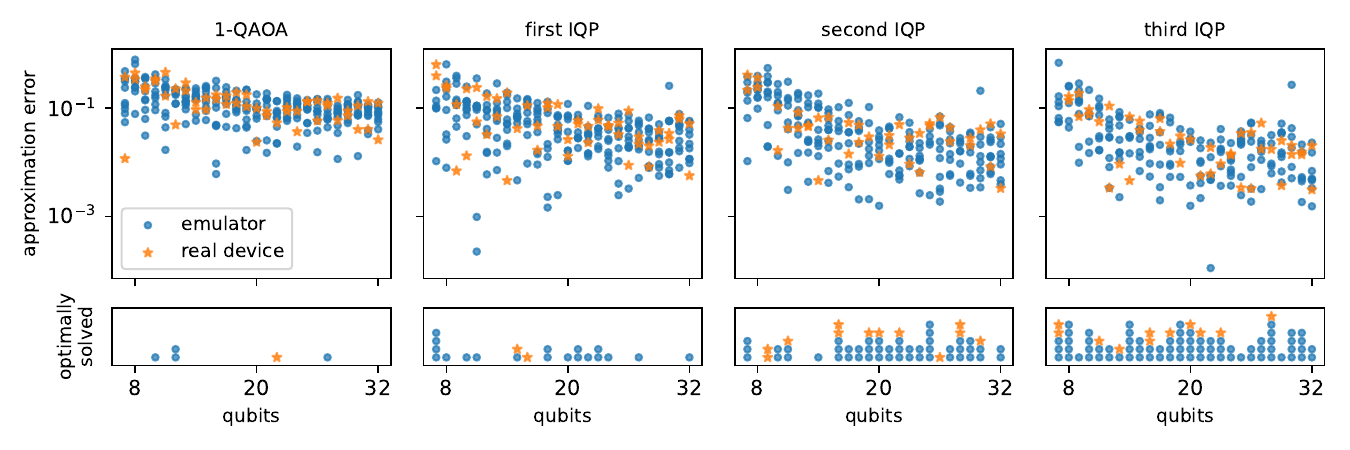}
    \caption{Optimization results on the Quantinuum H2 trapped-ion quantum hardware and emulator for randomly generated biased Sherrington-Kirkpatrick Hamiltonians of 7 to 32 qubits: two instances per problem size on the device (stars) and ten instances on the emulator (circles). For each instance we pick four steps along the gradient-descent trajectory, corresponding to the standard 1-layer QAOA, and three IQP circuits. Then take $\sim 2^{0.32N}\in[4,1208]$ shots, equally distributed across the four circuits. Each data point in the figure corresponds to the best solution sampled for each instance. If the best solution is optimal the point is placed in the lower row, while for sub-optimal solutions we place the point in the upper row to visualise the approximation error.}
        \label{fig:H2}
\end{figure*}

In Fig.~\ref{fig:solver}, we show how the overlap of the optimized IQP state onto the ground eigenspace varies with the problem size, and how it compares to the overlap achieved by the initial QAOA. Both plots show a clear exponential trend with relatively low and slowly increasing variance. This confirms that our algorithm has a significantly better exponential scaling compared to 1-layer QAOA. 

We also study how the temperature of the distribution changes as we perform imaginary time evolution on our variational manifold up to time $\tau = 10$, close to convergence. 
In Fig.~\ref{fig:temperature} we show how the optimal normalized temperatures achieved by the final optimized IQP state are lower than those achieved by the starting QAOA state. However, the KL divergence between the optimized IQP state and the best-fitting thermal state is higher and presents more dispersion than QAOA across Hamiltonian instances. This indicates that IQP states might be beneficial for the task of sampling low-energy eigenstates while QAOA provides a better approximation to thermal distributions.

In Fig.~\ref{fig:samples} we plot example distributions produced by the QAOA and the optimized IQP ansatz. From the qualitative aspect of the IQP distribution, we see that the performance of our algorithm in increasing the ground state overlap cannot be entirely explained as a consequence of having a lower temperature. The distribution becomes arched, and the probabilities of sampling the low-energy eigenstates rise orders of magnitude above the predictions of the thermal fit. Future theoretical work is necessary to understand how this effect emerges, and whether it is recovered in more general optimization problems.

Our algorithm is also studied in a more realistic setting, where quantum circuits are affected by hardware noise. We use the recently released Quantinuum H2 trapped-ion quantum hardware and emulator~\cite{Quantinuum}. The emulator performs exact state-vector simulation under a noise model that replicates the noisy behavior of the real device. The device presents all-to-all connectivity and high-fidelity parameterized gates of the form $\exp(-i\theta ZZ)$, making it ideal for our protocol and for QAOA on densely connected Hamiltonians.

For this analysis, we study biased SK models with the coefficients $h_i$ independently sampled from the same Gaussian distribution as the coefficients $J_{ij}$. The presence of the bias breaks the $\mathbb{Z}_2$ symmetry, halving the initial overlap with the ground eigenspace and making the problem slightly more general. The bias adds an additional slope in the vicinity of QAOA, that sometimes dissolves the local minimum that we exploit in the previous study, leaving no obvious method to pick a point in the gradient-descent trajectory.
Our aim for this analysis is however to study the performance in the neighborhood of QAOA, rather than providing the most optimized form of our protocol. For this purpose we pick the optimized 1-layer QAOA as the first circuit, and three equally-spaced circuits corresponding to three of the first gradient-descent steps. The Supplemental Material describes the criterion we used to pick these circuits.

Figure~\ref{fig:H2} compares the quality of the best solutions obtained by the corresponding four circuits.
From the 312 instances, we optimally solve 5, 21, 59, and 86, respectively for the four circuits. The best solution sampled for each instance has an average approximation ratio and standard deviation of $(0.87, 0.10)$, $(0.935, 0.083)$, $(0.948, 0.083)$, and $(0.970, 0.060)$, respectively. When considering for each instance only the best solution obtained from the four circuits as the output of our algorithm, $136$ instances are solved optimally (almost $44\%$) and the distribution has an average approximation ratio and standard deviation of $(0.985, 0.029)$.

\section{Discussion} \label{sec:discussion}
The algorithm we introduce explores the natural connection between the 1-layer QAOA state and IQP circuits. Studying the vicinity of the QAOA in this broader variational manifold leads to a better understanding of its optimality as a shallow-depth quantum heuristic, as well as how it can be improved. 

We show that, for the case of SK Hamiltonians, our approach amplifies the probability of sampling the ground state, beyond what can be obtained using classical tools such as MCMC. 
The hardware implementation is as resource-demanding as it is for 1-layer QAOA, and parameter training can be performed classically in time $\mathcal{O}(N^3)$. Results on the Quantinuum H2 show the reliability of our protocol to solve large instances with scarce quantum resources.

We leave as a future work the development of an optimized strategy to pick points along the gradient-descent trajectory where sampling from the quantum computer might yield even better performance.

The results presented motivate the development of strategies to compare the performance of our protocol against state-of-the-art classical algorithms at the scale of real-world combinatorial optimization problems. For example, the access to the analytical expectation value of the problem Hamiltonian and higher powers of it might provide an efficient way to estimate the probability of sampling the low-energy tail for large-scale problems.


\textit{Acknowledgments}---We thank Michael Lubasch, Luuk Coopmans, Marcello Benedetti, Matthias Rosenkranz, Matthew DeCross, Michael Foss-Feig and Cristina Cirstoiu for their feedback. We appreciate the helpful discussions with Christian Arenz and thank the authors of~\cite{Diez-Valle2023} for sharing their results with us. Finally, we acknowledge the great work of the Quantinuum H2 developing team: Steven Moses, Michael Mills, Jacob Johansen, Joan Dreiling, John Gaebler, Peter Siegfried, Juan Pino, and Caroline Figgatt.

\bibliography{bibliography}

\pagebreak

\onecolumngrid

In this supplemental material we present some technical aspects of the work that are not shown in the main text. Section \ref{app:analytic} contains a full derivation of the analytic expressions of expectation values of operators in the SD-IQP ansatz, as well as the application of this general formula to the particular case of a problem Hamiltonian. In Section \ref{app:counterexample} we prove that it is possible for an IQP optimization landscape to have local minima that do not correspond to eigenstates of the problem Hamiltonian, by constructing a minimal example. Section \ref{app:gram} shows how to obtain analytic expressions of the Gram matrix elements, which are essential to perform VarQITE. In Section \ref{app:landscape} we show examples of the cost function evolution until convergence and discuss the implications of the choice of sampling at different locations. Section \ref{app:criterion} discusses a criterion for selecting IQP circuits when the energy profile displays no plateaus.

\section{Analytic expression for expectation values in the IQP state} \label{app:analytic}
In this appendix, we derive exact analytic expressions of expectation values of low-weight Pauli operators in the IQP ansatz. A similar derivation is presented in~\cite{Hadfield2018} for 1-layer QAOA circuits, that contain only two parameters, so that resulting expressions are a particular case of the ones derived in this Appendix. In contrast to the expressions derived in~\cite{Herrman2022} for extra-parameterized 1-layer QAOA circuits, our expressions apply to arbitrary local Hamiltonians. During the write-up of the second version of this manuscript similar analytical expressions for arbitrary local Hamiltonians were obtained~\cite{Vijendran2023}.

Let $P_e$ be some Pauli string that applies non-identity Pauli operators to a subset of qubits $e \in \mathcal{N} = \{1, 2, \ldots, N\}$ and let $w_e = \abs{e}$ be the weight of $P_e$. If we identify Pauli strings that differ only by a phase, we can characterize them using two length-$N$ boolean vectors $a,b \in \mathbb{Z}_2^{\otimes N}$ by the decomposition $P_e = Z^aX^b$, where we used the notation $Z^a = \bigotimes_{i = 1}^N Z^{a_i}$. Denote by $e_Z$ and $e_X$ the subsets of $\mathcal{N}$ corresponding to the nonzero elements of $a$ and $b$ respectively, and let $w_a = \abs{e_Z}$ and $w_b = \abs{e_X}$ be corresponding weights. Our goal is then to compute $\langle Z^a X^b \rangle_{\boldsymbol{\theta}}$. First, we show how the layer of single qubit X rotations transforms this operator

\begin{equation}
\begin{split}
    \bigotimes_{i\in \mathcal{N}} R_x^{\dagger}(\phi_i) Z^a X^b \bigotimes_{i\in \mathcal{N}} R_x(\phi_i) &= \prod_{i\in \mathcal{N}} \left(R_x^{\dagger}(\phi_i) Z R_x(\phi_i)\right)^{a_i} \left(R_x^{\dagger}(\phi_i) X R_x(\phi_i)\right)^{b_i} \\
    &= \prod_{i\in \mathcal{N}} \left(\cos \phi_i Z + \sin \phi_i Y\right)^{a_i}  X^{b_i}.
\end{split}
\end{equation}

This can be expanded to a sum of $2^{w_a}$ Pauli strings, whose expectation values are then to be calculated in the state $\exp(-i \mathcal{H}_{\mathrm{IQP}})\ket{+}^{\otimes N}$. To simplify notation we will denote all expectation values in this state by $\langle \cdot \rangle$, which differs from the expectation value in the full ansatz by omitting the subscript $\boldsymbol{\theta}$. If we recycle previous notation for brevity, we are now interested in computing expectation values of the form
\begin{equation}
    \langle Z^a X^b \rangle =\bra{+}^{\otimes N} \exp(i \mathcal{H}_{\mathrm{IQP}}) Z^a X^b \exp(-i \mathcal{H}_{\mathrm{IQP}})\ket{+}^{\otimes N}.
\end{equation}

Since we are only working with IQPs we can expand the Hamiltonian as
\begin{equation}
    \mathcal{H}_{\mathrm{IQP}} = \frac{1}{2}\sum_{i\in \mathcal{N}} \theta_i Z_i+ \frac{1}{2}\sum_{i<j} \theta_{ij} Z_i Z_j.
\end{equation}

The terms in the Hamiltonian that commute with $Z^a X^b$ can be straight-forwardly canceled out, while those that anti-commute with $Z^a X^b$ can be moved through with a flipped sign. Then we have

\begin{equation}
\begin{split}
    \langle Z^a X^b \rangle &= \bra{+}^{\otimes N} Z^a \exp(i\sum_{i\in e_X} \theta_i Z_i + i\sum_{\substack{i\in e_X \\ j\notin e_X}}\theta_{ij}Z_i Z_j) X^b \ket{+}^{\otimes N} \\
    &= \frac{1}{2^N} \sum_{\{x\}} \bra{x} Z^a \exp(i\sum_{i\in e_X} \theta_i Z_i + i\sum_{\substack{i\in e_X \\ j\notin e_X}}\theta_{ij}Z_i Z_j) \ket{x},
\end{split}
\end{equation}
where in the last equality we expanded the state $\ket{+}^{\otimes N}$ as a sum over all spin configurations $x_i \in \{+1,-1\}$ and made use of the fact that the central operator is manifestly diagonal in this basis. We can absorb the $Z^a$ in the propagator by noting that $\exp(-i\pi Z/2) = -i Z$ and using the transformed angles $\tilde{\theta}_i = \theta_i - a_i\pi/2$, giving us the simple expression

\begin{equation}
    \langle Z^a X^b \rangle = \frac{i^{w_{a}}}{2^N} \sum_{\{x\}} \exp(i\sum_{i\in e_X} \tilde{\theta}_i x_i + i\sum_{j \notin e_X}x_j \left(-\frac{\pi}{2}a_j+\sum_{i \in e_X}\theta_{ij} x_i\right)).
\end{equation}

This form has the interpretation of a partition function over the bipartite graph formed by splitting the set of all qubits $\mathcal{N}$ into $e_X$ and its complement. This suggests we should separate the spins corresponding to different subsets, so we will denote by $r$ the configurations of spins in $e_X$ and by $s$ configurations of the complement. The expression is then rewritten as

\begin{equation}
\begin{split}
    \langle Z^a X^b \rangle &= \frac{i^{w_{a}}}{2^N} \sum_{\{s,r\}} \exp(i\sum_{i\in e_X} \tilde{\theta}_i r_i + i\sum_{j \notin e_X}s_j\left(-\frac{\pi}{2}a_j+\sum_{i \in e_X}\theta_{ij} r_i\right)) \\
    &= \frac{i^{w_{a}}}{2^N} \sum_{\{r\}} \exp(i\sum_{i\in e_X} \tilde{\theta}_i r_i ) \sum_{\{s\}} \exp( i\sum_{j \notin e_X}s_j\left(-\frac{\pi}{2}a_j+\sum_{i \in e_X}\theta_{ij} r_i\right)) \\ 
    &= \frac{i^{w_{a}}}{2^N} \sum_{\{r\}} \exp(i\sum_{i\in e_X} \tilde{\theta}_i r_i ) \prod_{j\notin e_X} \sum_{y \in \{+1,-1\}} (-iy)^{a_j} \exp( i\sum_{i \in e_X}\theta_{ij}r_i y) \\
    &= \frac{i^{w_{a}}}{2^{w_b}} \sum_{\{r\}} \exp(i\sum_{i\in e_X} \tilde{\theta}_i r_i ) \prod_{j\notin e_X} Q_{a_j}\left(\sum_{i \in e_X}\theta_{ij}r_i\right),
\end{split}
\end{equation}
which is now a sum over the configurations of spins in $e_X$ only. For simplified notation, we introduced the $Q$ function, which is defined as

\begin{align}
Q_0(x) &= \cos x, \\
Q_1(x) &= \sin x
\end{align}

We can simplify this even further by grouping configurations that differ only by the $\mathbb{Z}_2$ operation of flipping the sign of all spins to obtain

\begin{equation}
\begin{split}
    \langle Z^a X^b \rangle &= \frac{i^{w_{a}}}{2^{w_b}} \sum_{\{r\}/ \mathbb{Z}_2}\left[ \exp(i\sum_{i\in e_X} \tilde{\theta}_i r_i ) \prod_{j\notin e_X} Q_{a_j}\left(\sum_{i \in e_X}\theta_{ij}r_i\right) + \exp(-i\sum_{i\in e_X} \tilde{\theta}_i r_i ) \prod_{j\notin e_X} Q_{a_j}\left(-\sum_{i \in e_X}\theta_{ij}r_i\right)\right]\\  
    &= \frac{i^{w_{a}}}{2^{w_b}} \sum_{\{r\}/ \mathbb{Z}_2}\left[ \exp(i\sum_{i\in e_X} \tilde{\theta}_i r_i ) \prod_{j\notin e_X} Q_{a_j}\left(\sum_{i \in e_X}\theta_{ij}r_i\right) + \exp(-i\sum_{i\in e_X} \tilde{\theta}_i r_i ) \prod_{j\notin e_X} (-1)^{a_j}Q_{a_j}\left(\sum_{i \in e_X}\theta_{ij}r_i\right)\right]\\
    &= \frac{i^{w_{a}}}{2^{w_b}} \sum_{\{r\}/ \mathbb{Z}_2}\left[ \exp(i\sum_{i\in e_X} \tilde{\theta}_i r_i ) \prod_{j\notin e_X} Q_{a_j}\left(\sum_{i \in e_X}\theta_{ij}r_i\right) + (-1)^{a_P}\exp(-i\sum_{i\in e_X} \tilde{\theta}_i r_i ) \prod_{j\notin e_X} Q_{a_j}\left(\sum_{i \in e_X}\theta_{ij}r_i\right)\right]\\
    &= \frac{i^{w_{a}}}{2^{w_b}} \sum_{\{r\}/ \mathbb{Z}_2} \left[\exp(i\sum_{i\in e_X} \tilde{\theta}_i r_i )+ (-1)^{a_P}\exp(-i\sum_{i\in e_X} \tilde{\theta}_i r_i )\right]\prod_{j\notin e_X} Q_{a_j}\left(\sum_{i \in e_X}\theta_{ij}r_i\right),
\end{split}
\end{equation}
where $a_P = \sum_{j \notin e_X} a_j \mod 2$. We can now merge the two complex phases to obtain the final expression

\begin{equation}
\label{eq:expvalue}
    \langle Z^a X^b \rangle =\frac{i^{w_{a}+a_P}}{2^{w_b-1}} \sum_{\{r\}/ \mathbb{Z}_2} Q_{a_P}\left(\sum_{i\in e_X} \overline{\theta}_i r_i \right) \prod_{j\notin e_X} Q_{a_j}\left(\sum_{i \in e_X}\theta_{ij}r_i\right).
\end{equation}

This expresses the expectation value as a sum of $2^{w_b-1}$ terms and can be computed efficiently when the operators we are interested in have small weight $w \ll N$. In particular, to perform the optimization of the ansatz as described in the main text, we only make use of this expression with $w_b$ up to 2. Note that when $w_b = 0$ we have a vanishing expectation value. 

We will now show how Eq.~\eqref{eq:expvalue} can be used to efficiently compute the expectation of the Hamiltonian in the IQP

\begin{equation}
\label{eq:hamiltexp}
\begin{split}
    \langle \mathcal{H} \rangle_{\boldsymbol{\theta}} &= \sum_i h_i \langle Z_i \rangle_{\boldsymbol{\theta}} + \sum_{i<j} J_{ij} \langle Z_i Z_j \rangle_{\boldsymbol{\theta}} \\
    &= \sum_i h_i (\cos \phi_i \langle Z_i\rangle -i \sin \phi_i \langle Z_i X_i \rangle)  \\&+ \sum_{i<j} J_{ij}(\cos\phi_i\cos\phi_j \langle Z_i Z_j \rangle
    -i \cos\phi_i\sin\phi_j \langle Z_i Z_j X_j \rangle -i \sin\phi_i\cos\phi_j \langle Z_i Z_j X_i \rangle - \sin\phi_i\sin\phi_j \langle Z_i Z_j X_i X_j \rangle).
\end{split}
\end{equation}

We may now use Eq.~\eqref{eq:expvalue} to expand each term in this expression. First, we note that expectation values with no $X$ operators simply vanish, so $\langle Z_i \rangle =0$ and $\langle Z_i Z_j \rangle = 0$. Then we can compute the remaining terms individually

\begin{align}
    \langle Z_i X_i \rangle &= i \sin \theta_i \prod_{l \neq i} \cos \theta_{il}, \\
    \langle Z_i Z_j X_i \rangle &= i \cos \theta_i \sin \theta_{ij} \prod_{l \neq i,j} \cos \theta_{li}, \\
    \langle Z_i Z_j X_j \rangle &= i \cos \theta_j \sin \theta_{ij} \prod_{l \neq i,j} \cos \theta_{lj}, \\
    \langle Z_i Z_j X_i X_j \rangle &= \frac{1}{2}\left[ \cos (\theta_{i}+\theta_{j}) \prod_{l \neq i,j} \cos (\theta_{li}+\theta_{lj})-\cos (\theta_{i}-\theta_{j}) \prod_{l \neq i,j} \cos (\theta_{li}-\theta_{lj})\right].
\end{align}

If we plug these expressions into Eq.~\eqref{eq:hamiltexp} we arrive at the final analytic form for our Hamiltonian expectation value

\begin{equation}
    \begin{split}
        \langle &\mathcal{H} \rangle_{\boldsymbol{\theta}} = \sum_i h_i \sin \phi_i \sin \theta_i \prod_{l \neq i} \cos \theta_{il} \\
        &+ \sum_{i<j} J_{ij}\left[\cos \phi_i \sin \phi_j \cos \theta_j \sin \theta_{ij} \prod_{l \neq i,j} \cos \theta_{lj} + \sin \phi_i \cos \phi_j \cos \theta_i \sin \theta_{ij} \prod_{l \neq i,j} \cos \theta_{li} \right. \\
        &-\left.\frac{1}{2}\sin \phi_i \sin \phi_j \left( \cos (\theta_{i}+\theta_{j}) \prod_{l \neq i,j} \cos (\theta_{li}+\theta_{lj})-\cos (\theta_{i}-\theta_{j}) \prod_{l \neq i,j} \cos (\theta_{li}-\theta_{lj}) \right) \right].
    \end{split}
\end{equation}

Note that the computational time of evaluating this expectation value, as well as the gradient in $\boldsymbol \theta$, is $\mathcal{O} (N^3)$, if the problem Hamiltonian has all to all connectivity, as is the case for the SK model. It can be reduced to $\mathcal{O} (D N^2)$ if our problem can be formulated on a graph whose degree is bounded by $D$. Analogous efficient expressions may be obtained for problem Hamiltonians that include many-body interactions, as long as the weights scale at most like $\mathcal{O} (\log N)$ with the problem size.

\section{Example of non-trivial minima of the optimization landscape}
\label{app:counterexample}

In this appendix, we give a minimal example of a problem for which the IQP ansatz leads to non-trivial local minima, where by non-trivial we mean that the state produced is not an eigenstate of the Hamiltonian. Additionally, the state is shown to have an overlap of 0.5 onto the degenerate ground eigenspace, so the problem solution can be recovered by sampling.

Consider the 4-qubit Hamiltonian

\begin{equation}
    H = Z_0\left(Z_1+Z_2+Z_3\right).
\end{equation}

We explore this Hamiltonian using the IQP ansatz given by all 1-body and 2-body operators in the $X$-basis
\begin{align}
    \mathcal{H}^X_{IQP} = \sum_{i=0}^3 \theta_i X_i + \sum_{i<j} \theta_{ij} X_i X_j \\
    \ket{\Psi_{IQP}} = \exp\left(-\frac{i}{2}\mathcal{H}_{IQP}\right)\ket{0},
\end{align}
which is equivalent to the IQP state defined in the main text.

We compute the expectation value $\langle Z_0 Z_1 \rangle$ in the IQP state. The other 2 terms in H are obtained by cyclic permutations in 1, 2, and 3.

\begin{equation}
\begin{split}
    \langle Z_0 Z_1 \rangle &= \bra{0}\exp\left(\frac{i}{2}H_{IQP}\right) Z_0 Z_1 \exp\left(-\frac{i}{2}H_{IQP}\right) \ket{0} \\
    &= \bra{0}\exp\left[-i(\theta_0 X_0+\theta_1 X_1 + \theta_{02}X_0 X_2 +\theta_{03}X_0 X_3 + \theta_{12}X_1 X_2 +\theta_{13}X_1 X_3)\right]\ket{0}.
\end{split}
\end{equation}

Since all terms in the exponential commute, we can expand each factor individually as

\begin{equation}
    \langle Z_0 Z_1 \rangle = \bra{0}(\cos \theta_0 - i\sin \theta_0 X_0)\cdot \dots (\cos \theta_{13} - i\sin \theta_{13} X_1 X_3)\ket{0}.
\end{equation}

The only terms that survive are those that multiply to identity. The 4 products with this property are $I$, $X_0 (X_0 X_3) (X_3 X_1) X_1$, $X_0 (X_0 X_3) (X_3 X_1) X_1$, $(X_0 X_2)(X_2 X_1)(X_1 X_3) (X_3 X_0)$. Then we have

\begin{equation}
\begin{split}
    \langle Z_0 Z_1 \rangle = \cos \theta_0 \cos \theta_1 \cos \theta_{02} \cos \theta_{03} \cos \theta_{12} \cos \theta_{13}+ \\
    + \sin \theta_0 \sin \theta_1 \cos \theta_{02} \sin \theta_{03} \cos \theta_{12} \sin \theta_{13} + \\
    + \sin \theta_0 \sin \theta_1 \sin \theta_{02} \cos \theta_{03} \sin \theta_{12} \cos \theta_{13} + \\
    + \cos \theta_0 \cos \theta_1 \sin \theta_{02} \sin \theta_{03} \sin \theta_{12} \sin \theta_{13}.
\end{split}
\end{equation}

We provide a symbolic implementation of this expression in Python \cite{Leontica2022}. Using symbolic differentiation, we show that the line given by equations $\theta_0 = \pi/2,\, \theta_1 = \pi/2,\, \theta_3 = \pi/2,\, \theta_{01} = \pi/2,\, \theta_{02} = \pi,\, \theta_{12} = 0,\, \theta_{03} = \pi/2,\, \theta_{13} = \pi/2, \theta_{23} = 0$ (note that $\theta_2$ is free and parameterizes the line) is a critical line of local minima. This is done by verifying that the gradient is 0 for all $\theta_2$ and the hessian is positive semi-definite, with a single null eigenvalue corresponding to the direction going along the curve (except at the isolated point $\theta_2 = \pi/2$ which we exclude from our analysis). In addition, the expected value of the Hamiltonian on this line is $\langle H \rangle = -2$, while it is easy to check that all eigenvalues of the Hamiltonian must be odd integers. Therefore, the states created by the ansatz with the specified parameters must be superpositions of eigenstates with different eigenvalues. We claim that this is sufficient to show that all points on the line are non-trivial local minima and give the following proof:

\textit{Proof.} Consider an optimization space parameterized by $(\phi,\Vec{\theta})$ variables, with the usual parameter range of $0$ to $2\pi$. Let us call the cost function defined on this space by $J(\phi,\Vec{\theta})$. Assume $J$ is infinitely smooth, so we can freely Taylor expand around all points (this can be verified from the analytic form of $J$ in terms of sums of products of smooth functions). Assume the line of critical points found in the counterexample is defined by the condition $\Vec{\theta} = \Vec{\theta}_0$ for some constant $\Vec{\theta}_0$. $\phi$ can then be considered a parameterization of the critical line (changing its value moves us along the line). On this line, we showed that the cost function is constant

\begin{equation}
    J(\phi,\Vec{\theta}_0) = -2,
\end{equation}
and the gradient is 0

\begin{equation}
\label{eq:grad}
    \grad J(\phi,\Vec{\theta}_0) = 0.
\end{equation}

At the Hessian level, we find that all second-order derivatives that contain $\phi$ are zero

\begin{align}
    \partial_{\phi} \partial_{\theta_i} J (\phi, \Vec{\theta}_0) = 0, \\
    \partial_{\phi}^2 J (\phi, \Vec{\theta}_0) = 0,
\end{align}
and the restriction of the Hessian to the $\Vec{\theta}$ subspace is positive definite on some interval $[\phi_<, \phi_>]$ with $0<\phi_<<\phi_><\pi/2$:

\begin{equation}
    v_i \partial_{\theta_i}\partial_{\theta_j} J (\phi, \Vec{\theta}_0) v_j > 0
\end{equation}
for all nonzero vectors $\Vec{v}$ and $\phi \in [\phi_<, \phi_>]$. Einstein summation convention is employed. In particular, for all unit vectors $\hat{n}$ we have that

\begin{equation}
    n_i \partial_{\theta_i}\partial_{\theta_j} J (\phi, \Vec{\theta}_0) n_j \geq \lambda_{min}(\phi)>0,
\end{equation}
with $\lambda_{min}(\phi)$ the smallest eigenvalue of the Hessian at $(\phi, \Vec{\theta}_0)$ when restricted to the $\theta_i$ variables.

Note that the role of $\phi$ is played by $\theta_2$ in our example, but we changed the name for brevity. Suppose we want to focus our attention on the point $(\phi_0, \Vec{\theta}_0)$ with $\phi_0 \in [\phi_<, \phi_>]$ and show that this is indeed a local minimum. The coordinates of every point in the vicinity of this critical point that does not lie on the critical line can be written as $(\phi_0+\epsilon \Delta \phi,\Vec{\theta}_0+ \epsilon \Delta\Vec{\theta})$, where $\Delta\Vec{\theta}$ is a unit vector and $\epsilon$ is some small quantity. We can write the value of the cost function at this point as

\begin{equation}
\begin{split}
    J(\phi_0+\epsilon \Delta \phi,\Vec{\theta}_0+ \epsilon \Delta\Vec{ \theta}) = J(\phi_0+\epsilon \Delta \phi,\Vec{\theta}_0) + \\ \frac{\epsilon^2}{2}\Delta \theta_i \Delta \theta_j \partial_{\theta_i}\partial_{\theta_j} J (\phi_0+\epsilon \Delta \phi, \Vec{\theta}_0) + R(\epsilon),
\end{split}
\end{equation}

where from the theory of Taylor series we know that $R(\epsilon)$ is continuous and

\begin{equation}
\lim_{\epsilon \to 0} \frac{R(\epsilon)}{\epsilon^2} = 0.
\end{equation}

The linear term in the equation above has been dropped due to Eq.~\eqref{eq:grad} stating that all points of $\Vec{\theta} = \Vec{\theta}_0$ have vanishing gradients. Since the cost function is constant under shifts in $\phi$ this is equivalent to

\begin{equation}
\begin{split}
    &\frac{J(\phi_0+\epsilon \Delta \phi,\Vec{\theta}_0+ \epsilon \Vec{\Delta \theta})- J(\phi_0,\Vec{\theta}_0)}{\epsilon^2} = \\
    = \frac{1}{2}&\Delta \theta_i \Delta \theta_j \partial_{\theta_i}\partial_{\theta_j} J (\phi_0+\epsilon \Delta \phi, \Vec{\theta}_0) + \frac{R(\epsilon)}{\epsilon^2}
    \geq \frac{1}{2}\lambda_{min}(\phi_0+\epsilon\Delta\phi) + \frac{R(\epsilon)}{\epsilon^2}.
\end{split}
\end{equation}

Since the minimum eigenvalue is strictly positive in the $\vec{\theta}$ subspace and $R(\epsilon)/\epsilon^2$ is a continuous function that decays to 0 for $\epsilon \to 0$ we conclude that there must exist some $\epsilon_0>0$ such that the RHS of the above inequality is strictly positive for all $\abs{\epsilon}<\epsilon_0$. Since this implies that the LHS must also be positive for all values of $\epsilon$ in this ball around $0$, then this proves that the value of the cost function in the neighborhood must be strictly larger than its value on the critical line. The above prescription can be applied to all perturbations around the critical point except those with $\Delta\Vec{\theta} = 0$, for which we know that the cost function must be constant. This proves that all points on the critical line represent local minima (similar to the minima of the Mexican hat potential).

\section{Derivation of the Gram matrix}
\label{app:gram}

In order to perform VarQITE we must compute the Gram matrix $A$ corresponding to the tangent vectors of our variational manifold. According to its definition in~\cite{Yuan2019} we have

\begin{equation}
    A_{\mu \nu} = \Re \left(\frac{\partial \bra{\Psi(\boldsymbol{\theta})}}{\partial \boldsymbol{\theta}_\mu} \frac{\partial \ket{\Psi(\boldsymbol{\theta})}}{\partial \boldsymbol{\theta}_\nu} \right),
\end{equation}
where we used greek indices $\mu, \nu$ that run over all variational parameters in the ansatz. We can now separate this matrix into several blocks based on the type of variational parameter

\begin{equation}
    A = \begin{bmatrix}
    A^{\phi \phi} & A^{\phi \theta} \\
    A^{\theta \phi} & A^{\theta \theta}
    \end{bmatrix}.
\end{equation}

From the definition given in the main text we can explicitly compute the tangent vectors as

\begin{align}
    \frac{\partial \ket{\Psi(\boldsymbol{\theta})}}{\partial \phi_k} = -\frac{i}{2}\bigotimes_{i\in \mathcal{N}} R_x(\phi_i) X_k \exp\left(-i \mathcal{H}_{\mathrm{IQP}}\right) \ket{+}^{\otimes N}, \\
    \frac{\partial \ket{\Psi(\boldsymbol{\theta})}}{\partial \theta_e} = -\frac{i}{2}\bigotimes_{i\in \mathcal{N}} R_x(\phi_i) Z_e \exp\left(-i \mathcal{H}_{\mathrm{IQP}}\right) \ket{+}^{\otimes N},
\end{align}
where $e$ is a subset of $\mathcal{N}$ present in the IQP ansatz. For the $\theta \theta$ part of the matrix we get

\begin{equation}
    A^{\theta \theta}_{ee'} = \frac{1}{4} \Re ( \langle Z_e Z_{e'} \rangle) = \frac{1}{4} \delta_{ee'},
\end{equation}
where we again employ the notation $\langle \cdot \rangle$, which stands for the expectation value in the state $\exp\left(-i \mathcal{H}_{\mathrm{IQP}}\right) \ket{+}^{\otimes N}$. This result states that varying the parameters $\theta_e$ one at a time with the same starting point on the manifold will take us along orthogonal directions. For the $\theta \phi$ part and $\abs{e} = 1$ we have

\begin{equation}
    A^{\theta \phi}_{ik} = \frac{1}{4} \Re ( \langle Z_i X_k \rangle ) = \left\{\begin{array}{lr}
        0, & \text{for } k = i\\
        -\frac{1}{4}\sin \theta_k \sin \theta_{ik} \prod_{l \neq i,k} \cos \theta_{lk}, & \text{for } k \neq i
        \end{array} \right.,
\end{equation}
and for $\abs{e} = 2$ we get

\begin{equation}
    A^{\theta \phi}_{(ij)k} = \frac{1}{4} \Re ( \langle Z_i Z_j X_k \rangle ) = \left\{\begin{array}{lr}
        0, & \text{for } k = i \text{ or } k=j\\
        -\frac{1}{4}\cos \theta_k \sin \theta_{ik} \sin \theta_{jk} \prod_{l \neq i,j,k} \cos \theta_{lk}, & \text{for } k \neq i,j
        \end{array} \right..
\end{equation}

Since $A$ must be hermitian, we can get the other off-diagonal block matrix as $A^{\phi \theta} = (A^{\theta \phi})^T$. Finally in the $\phi \phi$ sector we have

\begin{equation}
    A^{\phi \phi}_{kq} = \frac{1}{4} \Re ( \langle  X_k X_q \rangle ) = \left\{\begin{array}{lr}
        1/4, & \text{for } k=q\\
        \frac{1}{8} \sum_{s \in \{+1,-1\}}\cos (\theta_{i}+s\theta_{j}) \prod_{l \neq i,j} \cos (\theta_{li}+s\theta_{lj}) & \text{for } k \neq q
        \end{array} \right..
\end{equation}

\section{Energy landscape}
\label{app:landscape}

\begin{figure*}[b!]
     \subfigure[]{
         \includegraphics[width=0.485\textwidth]{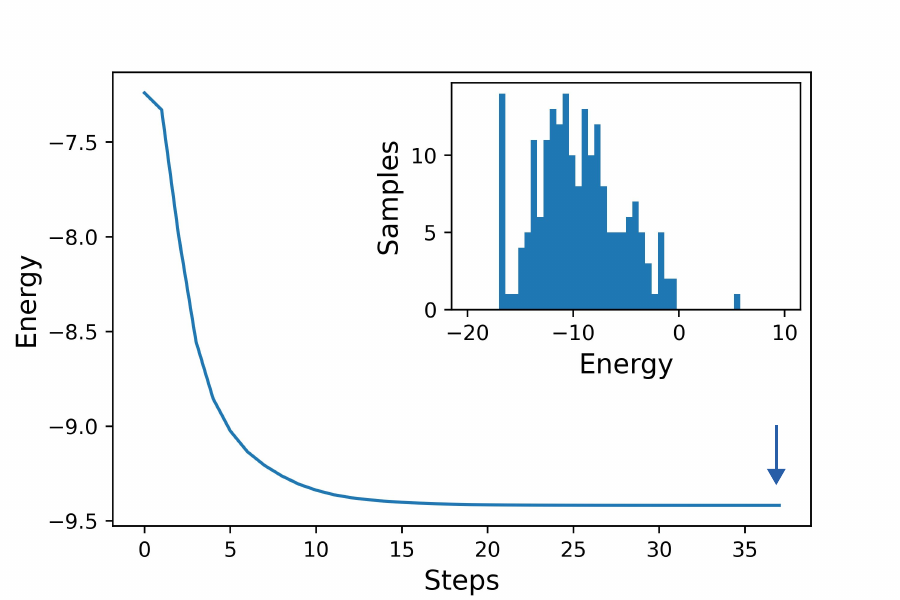}
     }
     \hfill
     \subfigure[]{
         \includegraphics[width=0.485\textwidth]{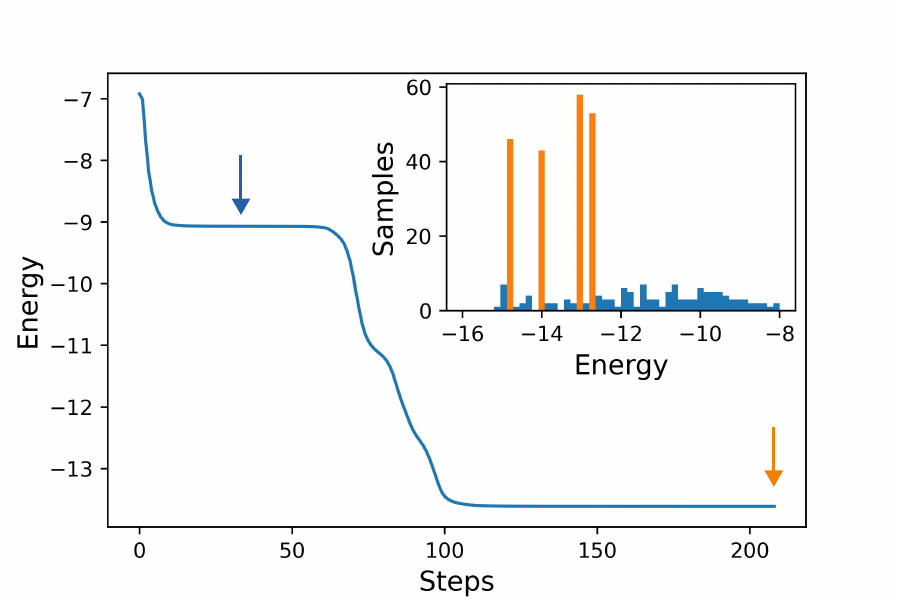}
         \label{fig:optbad}
     }
       \caption{Energy plot during simple gradient descent, starting from the optimal QAOA parameters for randomly generated 25 qubit optimization problems. a) Nearby non-trivial local minimum is found and the probability of sampling the ground state is amplified, b) Ansatz becomes degenerate after a long plateau. Orange shows samples collected from the final step and blue shows samples collected in the middle of the first plateau. All histograms are obtained from a total of 200 samples.}
        \label{fig:gradientdescent}
\end{figure*}

\begin{figure}[b!]
    \includegraphics[width=0.47\textwidth]{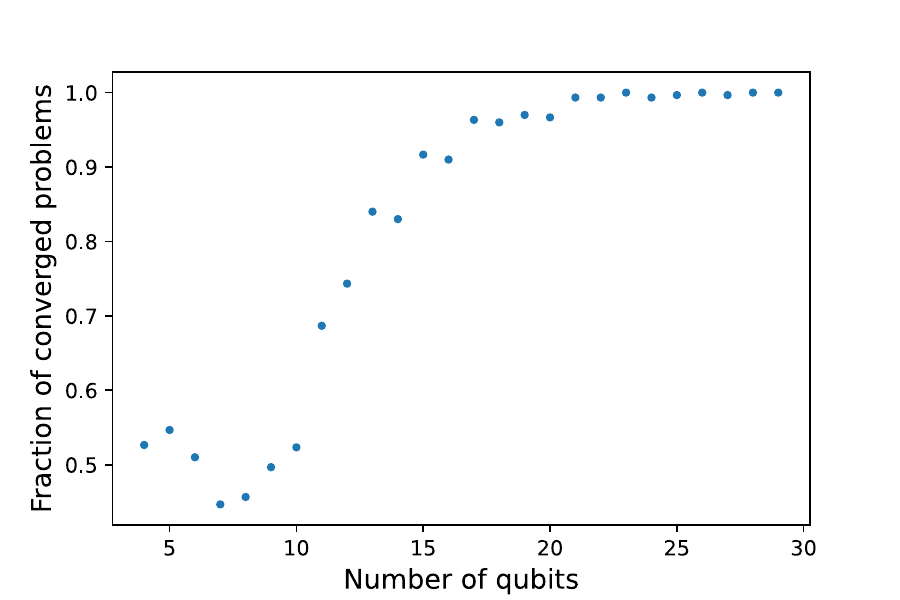}
    \caption{Fraction of problems that reach a local minimum near the starting QAOA position. The degenerate cases illustrated in Fig.~\ref{fig:optbad} become very rare as we increase the number of qubits.}
    \label{fig:probconv}
\end{figure}

In the main text, we claim that, for small problem sizes of the unbiased SK model, we sometimes do not find a non-trivial local minimum and instead the ansatz loses its overlap with the ground state. In this case, we still hit a plateau during the optimization and we overcome this issue by choosing to sample the IQP ansatz close to the middle of the plateau. This situation is illustrated in Fig.~\ref{fig:optbad}. In the attached histogram we see that the state obtained at the end of the optimization (orange) has non-zero support only on a small number of states and does not find the ground state, despite having much lower average energy. Sampling in the middle of the first plateau (blue) results in a wider spread of states, including a high overlap onto the ground state. In Fig.~\ref{fig:probconv} we show that the cases where the algorithm does not converge to a good local optimum become very unlikely as we increase the number of qubits.

\section{Criterion to select IQP circuits} \label{app:criterion}
In the case of biased SK Hamiltonians the local minimum in the vicinity of QAOA is usually dissolved by the bias and the average energy does not present an intermediate plateau. This leaves no obvious strategy to pick a point along the gradient-descent trajectory that provides some intuition on good performance. However, rather than optimizing this step of our protocol, the aim of this work is to study the performance in the neighborhood of QAOA. We then follow a simple criterion to pick four circuits in this vicinity.

The first circuit is precisely the optimized 1-layer QAOA, which serves as a warm-start to our protocol. To pick the other three circuits we observe the evolution of the parameters $\theta_{ij}$ in the two-qubit gates of the ansatz. When these parameters reach values close to $0$ or $\pi$ the implementation of the corresponding gate can be replaced by single-qubit gates: the identity or two single-qubit $\pi$-rotations, respectively. The remaining two-qubit gates define the connectivity graph of the ansatz at every step. 

The motivation to remove two-qubit gates with parameters close to $0$ or $\pi$ is that the noise inserted by their hardware implementation can be larger than the error incurred when they are replaced by their approximation as the identity or single-qubit gates, respectively. We decide to remove a two-qubit gate with parameter $\theta_{ij}$ close to $0$ (or $\pi$) if the effect of the gate is smaller than the reported infidelity $p \sim 10^{-3}$ in Quantinuum's devices, i.e., if $\abs{\sin(\theta_{ij}/2)} < p$ (or $\abs{\cos(\theta_{ij}/2)} < p$). 

Our criterion for a fair comparison to 1-layer QAOA is that this graph is at least connected, i.e.~formed by a single graph component. This ensures that, as for QAOA, the IQP circuit can not be split and implemented via separate unconnected circuits.

We pick the fourth circuit as the last circuit in the trajectory where the graph is still connected, and the second and third circuits at equidistant steps in that range. The fraction of entangling gates left in the three IQP circuits compared to QAOA has an average of $0.83$ and standard deviation across problem instances of $0.25$.

\end{document}